\documentclass[a4paper,12pt]{article}
\pdfoutput=1

\usepackage[pdftex]{graphicx}
\usepackage{subfigure}
\usepackage{float}
\usepackage{ textcomp }

\newcommand{\be}{\begin{equation}}
\newcommand{\ee}{\end{equation}}

\newcommand{\bea}{\begin{eqnarray}}
\newcommand{\eea}{\end{eqnarray}}

\newcommand{\p}{\partial}

\newcommand{\nn}{\nonumber \\}
\newcommand{\f}{\frac}
\newcommand{\w}{\wedge}
\newcommand{\ra}{\rightarrow}
\begin{document}
	\thispagestyle{empty}
	\begin{flushright}
		{\bf arXiv:1901.07175}
	\end{flushright}
	\begin{center} \noindent \Large \bf Thermal and thermoelectric conductivity of Einstein-DBI system
	\end{center}
	
	\bigskip\bigskip\bigskip
	\vskip 0.5cm
	\begin{center}
		{ \normalsize \bf   Shesansu Sekhar Pal
		}

		\vskip 0.5 cm

		 Department of Physics, Utkal University,  Bhubaneswar, 751004, India
		\vskip 0.5 cm
		\sf { shesansu${\frame{\shortstack{AT}}}$gmail.com 
		}
	\end{center}
	\centerline{\bf \small Abstract}
	
	We calculate the thermoelectric and thermal conductivity of Einstein-DBI system without the translational symmetry. We show that strictly, the longitudinal  component of the thermal conductivity in the vanishing dissipation limit, $k^2\rightarrow 0$,  is  determined only by the metric components evaluated at the horizon of the black hole in the same limit.  The temperature dependence of the transport quantities are obtained for $AdS_4$  and  $AdS_2\times R^2$ solutions.  The Lorentz ratio and Hall-Lorentz ratio   are found to be independent of temperature for  $AdS_4$ solution in the low temperature limit, whereas  the above mentioned quantities are independent of temperature for   $AdS_2\times R^2$ solution.

	\newpage
	\section{Introduction}
	
	After the advancement of gauge/gravity duality \cite{Maldacena:1997re}, there has been an  urge to understand the strongly coupled systems through it. In this context, several attempts are made to calculate the transport quantities in various  "theoretically  motivated" condensed matter systems, as summarized e.g., in \cite{Hartnoll:2016apf}, \cite{Ammon}, \cite{Zaanen} and \cite{Hartnoll:2007ih}. 
	
	In recent years, the computation of both the electrical conductivity  as well as the thermal conductivity  are studied  for  charged systems like Einstein-Maxwell system  as well as the Einstein-DBI system\footnote{Such a system is defined by having a space filling brane, whose action is described by the sum of Einstein and DBI action. Solutions and some other properties are studied e.g., in \cite{Pal:2010sx, Tarrio:2013tta, Kundu:2016oxg, Gushterov:2018spg,Pal:2017xpg, Gushterov:2018nht},}  e.g., in \cite{Karch:2007pd, OBannon:2007cex, Iqbal:2008by, Hartnoll:2008kx, Hartnoll:2009ns,  Hartnoll:2007ip,Davison:2013jba, Blake:2013bqa, Pal:2012zn, Kovtun:2008kx, Pal:2010sx, Charmousis:2010zz, Lee:2011qu, Chen:2017gsl, Wu:2017exh, Zhou:2015dha, Kim:2011zd,  Amoretti:2017axe, Davison:2015bea,Kiritsis:2016cpm} without momentum dissipation. The calculations are mostly performed following \cite{Iqbal:2008by}, where a flow equation of the electrical conductivity is obtained. Imposition of the in-falling  boundary condition at the horizon gives the desired electrical conductivity at the horizon. For the brane system,  the calculations are   performed following the recipe outlined in \cite{Karch:2007pd}.
	
	Recently, Donos and Gauntlett \cite{Donos:2014cya} and Gouteraux \cite{Gouteraux:2014hca}, separately  gave a prescription  to calculate the conductivities with momentum dissipation.  This prescription has been adopted to carry out the computation of the transport quantities for Einstein-Maxwell system with dissipation e.g., in  \cite{Blake:2014yla,Blake:2015ina, Kim:2015wba,Cremonini:2017qwq}

	In this paper, we shall report on the  calculation of the  magneto-electrical conductivities, magneto-thermoelectric and magneto-thermal conductivities for Einstein-DBI-dilaton-axion system without the translational symmetry. Moreover, it is done in the presence of charge density and  the case  for which the (fluctuated) electric field is perpendicular to the background magnetic field.

	The prescription for the calculation of the transport quantities 
	is to find  conserved quantities with respect to the radial flow. One radial flow is with respect to electrical current and the other is with respect to heat current. Then use of  ohm's law gives us the desired quantities for the electrical conductivities and the use of Fourier's law gives us the thermal conductivities,  which is usually the case in gauge/gravity duality. 
	
	We would like to mention that   with the result of \cite{Donos:2014cya}, it follows that in the absence of magnetic field, the metric component $h_{tx}$ diverges as the dissipation vanishes, $k^2\ra 0$, in which case, the linear order in perturbation breaks down. Hence, in such a scenario the zero dissipation limit is very subtle. However, in the presence of  a magnetic field, none of the fluctuating metric components diverges and in fact one can study the perturbation theory to any order, consistently. So, it is possible to consider the zero dissipation   limit consistently in the presence of a magnetic field.  
	
	For the Einstein-DBI-dilaton-axion system the calculation of the longitudinal electrical conductivity, longitudinal thermal conductivity and the longitudinal thermoelectric conductivity with dissipation is   reported in \cite{Blake:2017qgd} and the Hall electrical conductivity reported in \cite{Cremonini:2017qwq}. In this paper,  we report all the transport quantities  including the transverse thermal conductivity   as well as the transverse thermoelectric conductivity, which are new. Moreover, we find  the precise temperature dependence  of these transport quantities by studying  two explicit examples, one  at UV and the other at  IR.
	
It is suggested  that in the presence of a finite charge density and without magnetic field, if we fluctuate only one spatial component of the gauge field and one component of metric then the resulting longitudinal electrical conductivity can be divided into two pieces: one piece without having any dependence on the charge density and the other piece  with dependence on charge density and momentum dissipation as in  \cite{Donos:2014cya}. In \cite{Blake:2014yla}, the authors have interpreted the part without the charge density that contributes to the longitudinal conductivity is coming because of the charge conjugation symmetry, which means it respects the charge conjugation symmetry and arises  due to the particle hole pair.  The other part  is due to  momentum   dissipation. 

However, if we fluctuate along more than one component of the gauge field and the metric component in the presence of a background magnetic field  then  by considering a particular example of Einstein-Maxwell system as observed  in \cite{Blake:2015ina} and \cite{Kim:2015wba} that such a decomposition of the longitudinal electrical conductivity into two parts is not possible. Here we  demonstrate that such a decomposition into two parts is not possible for Einstein-DBI system too. We provide an interpretation of this as follows. Due to the infalling boundary condition at the horizon the black hole produces   current. The total current at the horizon is the  sum of currents due to the electric fields or the particle hole pairs in the language of \cite{Donos:2014cya, Blake:2014yla} and due to the metric fluctuations. In  the presence of dissipation the current due to metric fluctuations  gets mixed up with the current produced by the particle hole pair. Hence, the longitudinal part of the conductivity depends on the dissipation in a very non-trivial way.

	Another interesting result that follows from our study  is that the longitudinal thermal conductivity, $\kappa_{xx}$, in the zero dissipation limit can be expressed completely in terms of the metric components calculated at the horizon of the black hole in that limit, see eq(\ref{thermal_cond_horizon_zero_k}). In \cite{Blake:2017qgd}, the authors considered a quantity, $\kappa_L\neq\kappa_{xx}$, which can be expressed completely in terms of the metric components for any value of the dissipation when evaluated  at the horizon of the black hole.
	
	We have solved the Einstein-DBI-dilaton-axion system and obtained an exact  solution at UV, i.e.,  the $AdS_4$ solution. For such a solution, the longitudinal electrical conductivity, the Hall conductivity  as well as the longitudinal thermoelectric conductivity and the transverse  thermoelectric conductivity are found to be independent of temperature in the limit of vanishing dissipation and at zero temperature. However, the longitudinal thermal conductivity and the transverse thermal conductivity depends linearly on the temperature in the limit of vanishing dissipation and at low temperature. The ratio of longitudinal component of thermal conductivity to the temperature times the longitudinal component of electrical conductivity, $\frac{\kappa_{xx}}{T_H\sigma_{xx}}$,  and the ratio of transverse component of thermal conductivity to the temperature times the transverse component of electrical conductivity, $\frac{\kappa_{xy}}{T_H\sigma_{xy}}$, are independent of temperature in the zero temperature limit.
	
We have also obtained a solution at IR
 and our study reveals that at IR, i.e., for  the $AdS_2\times R^2$ solution,  the longitudinal as well as the transverse  component of the  thermal conductivities depends linearly on the temperature. However, the  electrical conductivities and the thermoelectric  conductivities are found to be independent of temperature. For this solution the ratio of longitudinal(transverse) component of thermal conductivity to the temperature times the longitudinal(transverse) component of electrical conductivity, i.e.,  $\frac{\kappa_{xx}}{T_H\sigma_{xx}}$, and $\frac{\kappa_{xy}}{T_H\sigma_{xy}}$ are found to be independent of temperature.

	The plan of the paper is as follows: in {\bf section 2}, we shall introduce our system  and obtain the necessary equations that the fields should satisfy. In {\bf section 3}, we shall include the fluctuations, obtain the equations obeyed at the linear order in fluctuation. Then obtain the  conserved quantities for the  electrical as well as the heat currents. We shall evaluate these currents at the horizon with the in-falling boundary conditions at the horizon. In {\bf section 4}, we shall express the currents in terms of the electric field as well as the thermal gradient, from which, we shall read out the desired conductivities. In {\bf section 5 and 6}, the $AdS_4$ solution as well as the $AdS_2\times R^2$ solution shall be obtained by solving the necessary equations and study the temperature dependence  of the transport quantities. Finally, we conclude in {\bf section 7}.

	\section{The system}
	
	The  low energy effective action that we shall be considering  has the following form\footnote{Upon doing the following transformation: $Z_1(\phi)\ra Z_1(\phi)Z^2_2(\phi) $ and 
	$Z_2(\phi)\ra Z^{-1}_2(\phi) $ with  the choice of setting the tension of brane to unity, we can reproduce the action considered in \cite{Blake:2017qgd}.}
	\bea\label{action}
	S&=&\f{1}{2\kappa^2}\int d^{4}x \Bigg[\sqrt{-g}\bigg({\cal R}-2\Lambda-\f{1}{2}(\p \phi)^2-V(\phi)-\f{W_1(\phi)}{2}(\p \chi_1)^2-\f{W_2(\phi)}{2}(\p \chi_2)^2\bigg)\nn &-&T_b Z_1(\phi)\sqrt{-det\bigg( Z_2(\phi) g_{MN}+\lambda F_{MN}\bigg)}\Bigg],
	\eea
where $Z_1~Z_2,~W_1,~W_2$ and $V$ are unknown functions of the scalar field, $\phi$. It describes the dynamics involving metric, scalar field, $\phi$, pseudoscalar fields, $\chi_i$ and the gauge field, $A_{M}$ and  $\lambda$ is a dimensionful constant.

The equation of motion of the metric components that follows from it takes the following form
\bea
&&R_{MN}-\f{1}{2} g_{MN}R+\f{1}{2}g_{MN}\bigg[2\Lambda+\f{1}{2}(\p \phi)^2+V(\phi)+\f{W_1(\phi)}{2}(\p \chi_1)^2+\f{W_2(\phi)}{2}(\p \chi_2)^2 \bigg]\nn&-&\f{1}{2}\p_M\phi\p_N\phi-\f{W_1(\phi)}{2}\p_M\chi_1\p_N\chi_1-
\f{W_2(\phi)}{2}\p_M\chi_2\p_N\chi_2+\f{T_b}{4}Z_1(\phi)Z_2(\phi)\times\nn&&\f{\sqrt{-det\bigg(Z_2(\phi)g +\lambda F\bigg)_{PS}}}{\sqrt{-g}}\bigg[\bigg(Z_2(\phi)g+\lambda F\bigg)^{-1}+{\bigg(Z_2(\phi)g~ -\lambda F\bigg)}^{-1} \bigg]^{KL}
g_{MK}g_{NL} =0\nn
\eea

The equation of motion of the gauge field is
\be\label{eom_gauge}
\p_M\Bigg[Z_1(\phi)\sqrt{-det\bigg(Z_2(\phi)g+\lambda F\bigg)_{PS}}\Bigg(\bigg(Z_2(\phi)g+\lambda F\bigg)^{-1}-{\bigg(Z_2(\phi)g-\lambda F\bigg)}^{-1}  \Bigg)^{MN}\Bigg]=0, 
\ee
where $F_{MN}=\p_M A_N-\p_N A_M$. The equations of motion of the scalar field, $\phi$, is
\bea\label{scalar_eom}
&&\p_{M}\bigg(\sqrt{-g}\p^M\phi \bigg)-\sqrt{-g}\bigg[\f{dV(\phi)}{d\phi}+\f{1}{2}\f{dW_1(\phi)}{d\phi}(\p \chi_1)^2+\f{1}{2}\f{dW_2(\phi)}{d\phi}(\p \chi_2)^2\bigg]-\nn&&T_b \f{dZ_1(\phi)}{d\phi}\sqrt{-det\bigg(g~ Z_2(\phi)+\lambda F\bigg)_{KL}}-\nn&&\f{T_b}{2}Z_1(\phi)
\sqrt{-det\bigg(g~ Z_2(\phi)+\lambda F\bigg)_{KL}}\f{dZ_2(\phi)}{d\phi}{\bigg(g~ Z_2(\phi)+\lambda F\bigg)}^{-1MN}g_{MN}=0.
\eea
The equation of motion associated to $\chi_1$ and $\chi_2$ are
\be
\p_{M}\bigg(\sqrt{-g}W_1(\phi)\p^M\chi_1 \bigg)=0,\quad\quad\quad \p_{M}\bigg(\sqrt{-g}W_2(\phi)\p^M\chi_2 \bigg)=0.
\ee

\paragraph {Ansatz:} For our purpose, we shall consider an ansatz where the metric is diagonal and there could exists  rotational symmetry along the $x-y$ plane involving the $ x_i$'s in the metric. The abelian field strength and the scalar field is assumed to be of the following form 
\bea\label{ansatz_solution}
ds^2_{3+1}&=&-U(r)dt^2+\f{1}{U(r)}dr^2+h(r)(dx^2+dy^2),\quad
\quad \phi=\phi(r),\nn
 \chi_1&=&k~~x, \quad \chi_2=k~~y,\quad A=A_t(r)dt+\f{B}{2}\left(-ydx+xdy \right),
\eea 
essentially, the metric components, scalar field and gauge fields are considered to be function of the radial coordinate only. The magnetic field, $B$, is constant. Importantly, the pseudo scalar field, $\chi_i$'s, breaks both the  translational as well as rotational symmetry.   For general solution with non-trivial value of the scalar field at IR and without the pseudo scalar field can be found in \cite{Pal:2012zn}. For simplicity, let us set $W_1(\phi)=W_2(\phi)=\psi(\phi)$

For such a choice of the ansatz the different metric components  satisfy the following differential equations
\bea\label{background_sol}
&&-4 \Lambda-2 V-\frac{2 k^2 \psi}{h}-\frac{2 T_b Z_1 {Z_2}^2 \left(B^2 \lambda ^2+h^2 {Z_2}^2\right)}{h \sqrt{\left(B^2 \lambda ^2+h^2 {Z_2}^2\right) \left({Z_2}^2-\lambda ^2 {A'_t}^2\right)}}-U \phi'^2-2 \f{h'U'}{h}+\nn &&\frac{U (h'^2-4 h h'')}{h^2}=0,
\nn
&& -\frac{U h'^2}{h}+2 h' U'+h \left(4 \Lambda +2 V+\frac{2 {T_b} h {Z_1}{Z_2}^2 \left({Z_2}^2-\lambda ^2{A'_t}^2\right)}{\sqrt{\left(B^2 \lambda ^2+h^2 {Z_2}^2\right) \left({Z_2}^2-\lambda ^2 {A'_t}^2\right)}}+U \phi'^2\right)+\nn&&2 U h''+2 h U''=0, \nn 
&& \frac{2 k^2 \psi}{h U}+\frac{2 {T_b} {Z_1} {Z_2}^2 \left(B^2 \lambda ^2+h^2 {Z_2}^2\right)}{h U \sqrt{\left(B^2 \lambda ^2+h^2 {Z_2}^2\right) \left({Z_2}^2-\lambda ^2 {A'_t}^2\right)}}+\frac{h' \left(U h'+2 h U'\right)}{h^2 U}+\phi'^2+\nn &&\frac{2 \left(2 \Lambda +V-U \phi'^2\right)}{U}=0,
\eea
where ${}^\prime$ denotes derivative with respect to $r$. The quantity $\sqrt{-det\bigg(Z_2(\phi)g+\lambda F\bigg)_{PS}}=\sqrt{Z^2_2(\phi)-A'^2_t}\sqrt{h^2Z^2_2(\phi)+B^2}$ for $\lambda=1$.
Since all the quantities in the equation of motion of the gauge field depends only on the radial direction means we should take derivative to act only  along  the radial direction. By considering $N=0$ in the equation of the gauge potential, eq(\ref{eom_gauge}), the equation reduces to 
$\p_r\Bigg[Z_1(\phi) \frac{A'_t\sqrt{Z^2_2(\phi)h^2+ B^2}}{\sqrt{Z^2_2(\phi)-A'^2_t}}\Bigg]=0$. Upon integrating it we get 
$A'_t=\frac{\rho Z_2(\phi)}{\sqrt{\rho^2+ Z^2_1(\phi)(B^2+Z^2_2(\phi)h^2)}}$, where the constant of integration $\rho$ is interpreted as the charge density.

The solution of the gauge potential can be obtained from 
\be
A'_t(r)=\f{\rho Z_2}{\sqrt{\rho^2+ B^2Z^2_1+Z^2_1Z^2_2h^2}}.
\ee

\section{Electrical and thermal Conductivity }

In order to calculate the conductivities, we need to find the conserved electrical currents and the heat currents. Then by taking the derivative of the appropriate currents with respect to the sources gives us the desired result. 

\subsection{Fluctuation and equations}

Let us fluctuate the background solution  following \cite{Donos:2014cya}:
\bea
ds^2&=&-U(r)dt^2+\f{1}{U(r)}dr^2+h(r)(dx^2+dy^2)+2G_{tx}(t, r)dtdx+
2G_{rx}(r)dxdr\nn&+&2G_{ty}(t, r)dtdy +2G_{ry}(r)drdy,\quad \chi_1=k~~x+\delta\chi_1(r), \quad \chi_2=k~~y+\delta\chi_2(r)\nn A&=&A_t(r)dt+\f{B}{2}\left(-ydx+xdy \right)+A_x(t,r)dx+A_y(t,r)dy,
\eea
where the fluctuating part of the metric and other fluctuating fields shall be considered infinitesimally. In particular, it has the following structure
\bea
A_x(t,r)&=&-E_x t+a_x(r)+\xi_x ~t~ A_t(r),\quad A_y(t,r)=-E_y t+a_y(r)+\xi_y ~t~ A_t(r)\nn
G_{tx}(t,r)&=&h(r) h_{tx}(r)-t~ \xi_x~  U(r),\quad G_{ty}(t,r)=h(r) h_{ty}(r)-t~ \xi_y ~ U(r),\nn
G_{rx}&=&h(r) h_{rx}(r),\quad G_{ry}=h(r) h_{ry}(r),
\eea
where $\xi_i$'s  are considered to be thermal gradients along the spatial directions.
With such a form of the geometry and matter field, the equation of motion associated to the metric components  with the choice $\lambda=1$ takes the following form
\bea\label{flu_eqs}
&&h U h''_{tx}+2h'U h'_{tx}-h_{tx}\left(\f{B^2 T_b  Z_1 {Z_2}^2}{ \sqrt{\left(B^2 +h^2 {Z_2}^2\right) \left({Z_2}^2- {A'_t}^2\right)}}+k^2\psi \right)+\nn &&T_b Z_1{Z_2}^2\left(\f{B(-E_y+\xi_y A_t)+hU A'_t(Bh_{ry}+a'_x)}{\sqrt{\left(B^2 +h^2 {Z_2}^2\right) \left({Z_2}^2- {A'_t}^2\right)}}\right)=0,\nn 
&&h U h''_{ty}+2h'U h'_{ty}-h_{ty}\left(\f{B^2 T_b  Z_1 {Z_2}^2}{ \sqrt{\left(B^2 +h^2 {Z_2}^2\right) \left({Z_2}^2- {A'_t}^2\right)}}+k^2\psi \right)+\nn &&T_b Z_1{Z_2}^2\left(\f{B(E_x-\xi_x A_t)+hU A'_t(-Bh_{rx}+a'_y)}{\sqrt{\left(B^2 +h^2 {Z_2}^2\right) \left({Z_2}^2- {A'_t}^2\right)}}\right)=0,\nn
&&2h U h' h_{rx}\left( B^2 T_b Z_1 {Z_2}^2+k^2\psi\sqrt{\left(B^2 +h^2 {Z_2}^2\right) \left({Z_2}^2- {A'_t}^2\right)}\right)-2B T_b h h' Z_1 {Z_2}^2Ua'_y\nn&&-2 k h U h'\psi\delta\chi'_1\sqrt{\left(B^2 +h^2 {Z_2}^2\right) \left({Z_2}^2- {A'_t}^2\right)}+2  T_b h^2 h' Z_1  {Z_2}^2 A'_t(E_x-B h_{ty})\nn &&+\xi_x  \sqrt{\left(B^2 +h^2 {Z_2}^2\right) \left({Z_2}^2- {A'_t}^2\right)}\Biggl( h^2 U \phi'^2-4h^2\Lambda-2h^2 V-2k^2 h \psi-3Uh'^2\nn &-& \f{2T_b hZ_1 {Z_2}^2 (B^2 +h^2 {Z_2}^2)}{\sqrt{\left(B^2 +h^2 {Z_2}^2\right) \left({Z_2}^2- {A'_t}^2\right)}}\Biggr)-2\xi_xT_b h^2 Z_1 {Z_2}^2 h'A_t A'_t=0,\nn
&&2h U h' h_{ry}\left( B^2 T_b Z_1 {Z_2}^2+k^2\psi\sqrt{\left(B^2 +h^2 {Z_2}^2\right) \left({Z_2}^2- {A'_t}^2\right)}\right)+2B T_b h h' Z_1 {Z_2}^2Ua'_x\nn&&-2 k h U h'\psi\delta\chi'_2\sqrt{\left(B^2 +h^2 {Z_2}^2\right) \left({Z_2}^2- {A'_t}^2\right)}+2  T_b h^2 h' Z_1  {Z_2}^2 A'_t(E_y+B h_{tx})\nn &&+\xi_y  \sqrt{\left(B^2 +h^2 {Z_2}^2\right) \left({Z_2}^2- {A'_t}^2\right)}\Biggl( h^2 U \phi'^2-4h^2\Lambda-2h^2 V-2k^2 h \psi-3Uh'^2\nn &-& \f{2T_b h Z_1 {Z_2}^2 (B^2 +h^2 {Z_2}^2)}{\sqrt{\left(B^2 +h^2 {Z_2}^2\right) \left({Z_2}^2- {A'_t}^2\right)}}\Biggr)-2\xi_y T_b h^2 Z_1 {Z_2}^2 h'A_t A'_t=0.
\eea

\subsection{Currents}
We shall calculate two types  currents, one associated to electrical type and the other associated to heat type. The convention that we shall be following is by setting $2\kappa^2=1$. 

\paragraph{Electrical currents:} The electrical currents are very easy to calculate and follows from the action as written down in eq(\ref{action}) which depends only on the field strength.  After the  inclusion of the fluctuations, it follows trivially that the action is independent of the gauge potentials, $a_x$ and $a_y$. Moreover, the fluctuation of the metric components do not depend on the gauge potentials,  $a_x$ and $a_y$ but rather on its derivative. 

It follows that the conserved currents along x-direction and y-direction are
\bea\label{electrical_current}
J^x(r)&=&\f{T_b Z_1 \left(BA'_tE_y-h^2 A'_t h_{tx}{Z_2}^2-hU{Z_2}^2(Bh_{ry}+a'_x)\right)}{\sqrt{\left(B^2 +h^2 {Z_2}^2\right) \left({Z_2}^2- {A'_t}^2\right)}}\nn
&-&\xi_y T_b B\left( \f{Z_1 A_t A'_t}{\sqrt{\left(B^2 +h^2 {Z_2}^2\right) \left({Z_2}^2- {A'_t}^2\right)}
} +\int^r_{r_h}dr\f{Z_1\left({Z_2}^2- {A'_t}^2\right)}{\sqrt{\left(B^2 +h^2 {Z_2}^2\right) \left({Z_2}^2- {A'_t}^2\right)}}\right),\nn
J^y(r)&=&-\f{T_b Z_1 \left(BA'_tE_x+h^2 A'_t h_{ty}{Z_2}^2+hU{Z_2}^2(-Bh_{rx}+a'_y)\right)}{\sqrt{\left(B^2 +h^2 {Z_2}^2\right) \left({Z_2}^2- {A'_t}^2\right)}}\nn
&+&\xi_x T_b B\left( \f{Z_1 A_t A'_t}{\sqrt{\left(B^2 +h^2 {Z_2}^2\right) \left({Z_2}^2- {A'_t}^2\right)}
} +\int^r_{r_h}dr\f{Z_1\left({Z_2}^2- {A'_t}^2\right)}{\sqrt{\left(B^2 +h^2 {Z_2}^2\right) \left({Z_2}^2- {A'_t}^2\right)}}\right).
\eea
With little bit of calculation, one can check that the electrical currents are conserved, namely, it  satisfies, $\p_r J^x=0$ and $\p_r J^y=0$, upon using the equation of motion associated to  the gauge potentials.

\paragraph{Heat currents:} Let us construct the conserved  heat currents associated to the metric fluctuation. In order to do so, let us start with the following choice  \cite{Donos:2014cya}
\be
Q^x\equiv U^2~ \p_r\left(\f{h~ h_{tx}}{U} \right)-A_t J^x.
\ee
The derivative of $Q^x$ using the conservation of the electric current, $\p_r J^x=0$,  gives
\be
\p_r Q^x=U\left(h'' h_{tx}+2 h'h'_{tx}+h h''_{tx} \right)-U'' h h_{tx}-A'_t J^x.
\ee
Using eq(\ref{background_sol}) and eq(\ref{flu_eqs}), the quantity, $\p_r Q^x$ becomes
\be
\p_r Q^x=\f{B T_b(E_y-\xi_y A_t) Z_1\left({Z_2}^2- {A'_t}^2\right)}{\sqrt{\left(B^2 +h^2 {Z_2}^2\right) \left({Z_2}^2- {A'_t}^2\right)}}+\xi_y A'_t\int^r_{r_h}\f{B T_b Z_1\left({Z_2}^2- {A'_t}^2\right)}{\sqrt{\left(B^2 +h^2 {Z_2}^2\right) \left({Z_2}^2- {A'_t}^2\right)}}\equiv {\cal M}^x(r).
\ee

If we construct the following quantity
\bea
{\cal Q}^x(r)&\equiv&Q^x(r)-\int^r_{r_h}dr\left[U\left(h'' h_{tx}+2 h'h'_{tx}+h h''_{tx} \right)-U'' h h_{tx}-A'_t J^x \right],\nn
&=&U^2~ \p_r\left(\f{h~ h_{tx}}{U} \right)-A_t J^x-\int^r_{r_h}\left[U\left(h'' h_{tx}+2 h'h'_{tx}+h h''_{tx} \right)-U'' h h_{tx}-A'_t J^x \right],
\eea
then it follows immediately
\be
\p_r {\cal Q}^x=0.
\ee

Explicitly, the form of the conserved heat current along x-direction takes the following form
\be\label{heat_current_qx}
{\cal Q}^x(r)=U^2~ \p_r\left(\f{h~ h_{tx}}{U} \right)-A_t J^x-\int^r_{r_h}dr' {\cal M}^x(r').
\ee

Similarly, defining $Q^y \equiv U^2~ \p_r\left(\f{h~ h_{ty}}{U} \right)-A_t J^y$, one gets
\be
\p_r Q^y=-\f{B T_b(E_x-\xi_x A_t) Z_1\left({Z_2}^2- {A'_t}^2\right)}{\sqrt{\left(B^2 +h^2 {Z_2}^2\right) \left({Z_2}^2- {A'_t}^2\right)}}-\xi_x A'_t\int^r_{r_h}\f{B T_b Z_1\left({Z_2}^2- {A'_t}^2\right)}{\sqrt{\left(B^2 +h^2 {Z_2}^2\right) \left({Z_2}^2- {A'_t}^2\right)}}\equiv {\cal M}^y(r).
\ee

So the conserved heat current along y-direction, $\p_r{\cal Q}^y(r) =0$, is 
\be\label{heat_current_qy}
{\cal Q}^y(r)=U^2~ \p_r\left(\f{h~ h_{ty}}{U} \right)-A_t J^y-\int^r_{r_h}dr' {\cal M}^y(r').
\ee

\subsection{At the horizon} 
We shall investigate the behavior of different fields and currents at the horizon, which in turn  allows us to compute the transport quantities at the horizon.
If we consider the background geometry to be a black hole then the function $U(r)$ has a zero at the horizon, $r_h$. Near to the horizon, it can be expanded as $U(r)=U_0 (r-r_h)+\cdots $, where $U_0$ is a constant, in which case, the Hawking temperature reads as $T_H=U_0/(4\pi)$.

\paragraph{In-falling boundary condition:} The boundary conditions for  different fields at the horizon of the black hole are considered as follows \cite{Donos:2014cya}:
\bea
a_x&=&-\f{E_x}{U_0}~Log(r-r_h)+{\cal O}(r-r_h),\quad a_y=-\f{E_y}{U_0}~Log(r-r_h)+{\cal O}(r-r_h),\nn
h_{tx}&=&U h_{rx}-\xi_x \left(\f{U}{h~ U_0}\right)~Log(r-r_h)+{\cal O}(r-r_h),\nn h_{ty}&=&Uh_{ry}-\xi_y \left(\f{U}{h~ U_0}\right)~Log(r-r_h)+{\cal O}(r-r_h).
\eea

\paragraph{Currents at the horizon:} Since the electrical currents eq(\ref{electrical_current}) and the heat currents eq(\ref{heat_current_qx}) and eq(\ref{heat_current_qy})   evaluated for any value of the radial coordinate, $r$, are conserved. It means we  can evaluate it at the horizon too.

The currents at the horizon takes the following form

\bea\label{electric_current_horizon}
J^x(r_h)&=&\left[\f{T_b Z_1 \left(BA'_tE_y-h^2 A'_t h_{tx}{Z_2}^2-h{Z_2}^2(Bh_{ty}-E_x)\right)}{\sqrt{\left(B^2 +h^2 {Z_2}^2\right) \left({Z_2}^2- {A'_t}^2\right)}}\right]_{r_h}\nn
J^y(r_h)&=&-\left[\f{T_b Z_1 \left(BA'_tE_x+h^2 A'_t h_{ty}{Z_2}^2-h{Z_2}^2(Bh_{tx}+E_y)\right)}{\sqrt{\left(B^2 +h^2 {Z_2}^2\right) \left({Z_2}^2- {A'_t}^2\right)}}\right]_{r_h},
\eea\bea\label{heat_current_horizon}
{\cal Q}^x(r_h)&=&\left[U^2~ \p_r\left(\f{h~ h_{tx}}{U} \right)-A_t J^x\right]_{r_h}=\left[U(h'h_{tx}+h h'_{tx})-U'h h_{tx}-A_t J^x\right]_{r_h},\nn
{\cal Q}^y(r_h)&=&\left[U^2~ \p_r\left(\f{h~ h_{ty}}{U}-A_t J^y \right)\right]_{r_h}=\left[U(h'h_{ty}+h h'_{ty})-U'h h_{ty}-A_t J^y\right]_{r_h}.
\eea

 Moreover, the gauge potential at the horizon vanishes, in which case, the heat currents become
\be\label{heat_current}
{\cal Q}^x(r_h)=-U'(r_h)h(r_h) h_{tx}(r_h),\quad\quad\quad {\cal Q}^y(r_h)=-U'(r_h)h(r_h) h_{ty}(r_h).
\ee

\paragraph{Behavior of $h_{tx}$ and $h_{ty}$ at the horizon:} In order to calculate the currents in terms of the electric fields and the thermal gradients, we need to know the value of the metric fluctuations,  $h_{tx}$ and $h_{ty}$, at the horizon. 

From the first two equations of  eq(\ref{flu_eqs}) obeyed by the metric components $h_{tx}$ and $h_{ty}$, it follows that  at the horizon 
\bea
&&h_{tx}\left[B^2 T_b \Omega +k^2 \psi \right]+B T_b \Omega E_y+ T_b h\Omega A'_t(E_x-B h_{ty)}+\xi_x U_0 =0,\nn
&&h_{ty}\left[B^2 T_b \Omega+k^2 \psi \right]+T_b \Omega h B A'_t  h_{tx}-T_b \Omega(BE_x-h A'_t E_y)+\xi_y U_0=0,
\eea
where $\Omega=Z_1 {Z^2_2}/\sqrt{\left(B^2 +h^2 {Z_2}^2\right) \left({Z_2}^2- {A'_t}^2\right)}$.

It is east to solve the coupled algebraic equations and the solution reads as
\bea\label{sol_htx_hty_horizon}
h_{tx}(r_h)&=&-\Biggl[\f{1}{[(B^2 T_b\Omega+k^2\psi)^2+B^2 T^2_b h^2 \Omega^2 A'^2_t]}\Biggl(E_xk^2 T_b\Omega h \psi A'_t+\xi_x U_0\left(B^2 T_b\Omega+k^2\psi\right)+\nn&&E_y BT_b \Omega\left(B^2 T_b\Omega+k^2\psi+T_b\Omega h^2 A'^2_t\right)+\xi_y BT_b U_0\Omega h A'_t \Biggr)\Biggr]_{r_h}, \nn 
h_{ty}(r_h)&=&\Biggl[\f{1}{[B^2 T_b\Omega+k^2\psi)^2+B^2 T^2_b h^2 \Omega^2 A'^2_t]}\Biggl( E_x B T_b\Omega\left(B^2 T_b\Omega +k^2 \psi+ T_b \Omega h^2 A'^2_t\right)-\nn&&E_y k^2 T_b\Omega h \psi A'_t+\xi_x BT_b U_0\Omega h A'_t-\xi_y U_0\left(B^2 T_b\Omega+k^2\psi \right)\Biggr)\Biggr]_{r_h}.
\eea

It is easy to notice that in the zero momentum dissipation limit, the fluctuations of the metric components, $h_{tx}$ and $h_{ty}$ are finite at the horizon. 

\section{Currents in terms of electric fields and thermal gradients}

\paragraph{Electrical current and heat current:} Upon substituting the solution of $h_{tx}(r_h)$ and $h_{ty}(r_h)$ from eq(\ref{sol_htx_hty_horizon}) in the expression of the currents evaluated at the horizon, eq(\ref{electric_current_horizon}) and eq(\ref{heat_current}), we get
\bea\label{electrical_heat_current}
J^x(r_h)&=&\sigma_{xx}(r_h)E_x+\sigma_{xy}(r_h)E_y+T_H\alpha_{xx}(r_h)\xi_x+T_H\alpha_{xy}(r_h)\xi_y,\nn
J^y(r_h)&=&\sigma_{yx}(r_h)E_x+\sigma_{yy}(r_h)E_y+T_H\alpha_{yx}(r_h)\xi_x+T_H\alpha_{yy}(r_h)\xi_y,\nn
{\cal Q}^x(r_h)&=&T_H{\overline\alpha}_{xx}(r_h)E_x+T_H{\overline\alpha}_{xy}(r_h)E_y+T_H{\overline\kappa}_{xx}(r_h)\xi_x+T_H{\overline\kappa}_{xy}(r_h)\xi_y,\nn
{\cal Q}^y(r_h)&=&T_H{\overline\alpha}_{yx}(r_h)E_x+T_H{\overline\alpha}_{yy}(r_h)E_y+T_H{\overline\kappa}_{yx}(r_h)\xi_x+T_H{\overline\kappa}_{yy}(r_h)\xi_y.
\eea
This definition of conductivities can be related to \cite{ Hartnoll:2007ih} by identifying the parameter $\xi_i$ with the temperature gradient as follows $T_H \xi_i=-\p_i T_H$. 
The electrical conductivities are defined as the response of the electrical currents to the electric fields and  reads as

\bea
\sigma_{xx}(r_h)&=&\sigma_{yy}(r_h)=\Biggl[\f{k^2 T_b\Omega h \psi \left(k^2\psi+T_b\Omega(B^2+h^2 A'^2_t)\right)}{(B^2 T_b\Omega+k^2\psi)^2+B^2 T^2_b\Omega^2 h^2 A'^2_t}\Biggr]_{r_h},
\eea
\bea
\sigma_{xy}(r_h)&=&-\sigma_{yx}(r_h)=\bigg[ \f{B T_b\Omega A'_t}
{Z^2_2((B^2 T_b\Omega+k^2\psi)^2+B^2 T^2_b\Omega^2 h^2 A'^2_t)} \bigg((B^2 T_b\Omega+k^2\psi)^2\nn&+&T_b\Omega h^2\left[2k^2 Z^2_2\psi+T_b\Omega(B^2(Z^2_2+A'^2_t)+Z^2_2h^2A'^2_t)\right] \bigg)\Biggr]_{r_h}.\nn
\eea
The thermoelectric conductivities are defined as either  the response of the electric currents to the thermal gradients or the  response of the heat currents to the electric fields and  are  
\bea
\alpha_{xx}(r_h)&=&\alpha_{yy}(r_h)={\overline\alpha}_{xx}={\overline\alpha}_{yy}=\f{1}{T_H}\Biggl[\f{k^2 T_b\Omega U_0 h^2 \psi A'_t}{(B^2 T_b\Omega+k^2\psi)^2+B^2 T^2_b\Omega^2 h^2 A'^2_t}\Biggr]_{r_h},
\eea
\bea
\alpha_{xy}(r_h)&=&-\alpha_{yx}(r_h)={\overline\alpha}_{xy}=-{\overline\alpha}_{yx}=\f{1}{T_H}\Biggl[\f{B T_b\Omega U_0 h\bigg(k^2 \psi+T_b\Omega(B^2+h^2 A'^2_t)\bigg)}{(B^2 T_b\Omega+k^2\psi)^2+B^2 T^2_b\Omega^2 h^2 A'^2_t}\Biggr]_{r_h}.
\eea

The thermal conductivity is defined as the response of the heat current  to the  thermal gradient for zero electric current and reads as,  $\kappa_{ij}={\overline\kappa}_{ij}-T_H(\alpha\sigma^{-1}\alpha)_{ij}$, with 
\bea
{\overline\kappa}_{xx}(r_h)&=&{\overline\kappa}_{yy}(r_h)=\f{1}{T_H}\Biggl[\f{ U^2_0 h (B^2T_b\Omega+k^2\psi)}{(B^2 T_b\Omega+k^2\psi)^2+B^2 T^2_b\Omega^2 h^2 A'^2_t}\Biggr]_{r_h},
\eea
\bea
{\overline\kappa}_{xy}(r_h)&=&-{\overline\kappa}_{yx}(r_h)=\f{1}{T_H}\Biggl[\f{B T_b\Omega U^2_0 h^2 A'_t}{(B^2 T_b\Omega+k^2\psi)^2+B^2 T^2_b\Omega^2 h^2 A'^2_t}\Biggr]_{r_h}.
\eea
This gives thermal conductivity as
\bea
\kappa_{xx}(r_h)&=&\kappa_{yy}(r_h)=\Biggl[{\overline\kappa}_{xx}-T_H\f{\left((\alpha^2_{xx}-\alpha^2_{xy})\sigma_{xx}+2\alpha_{xx}\alpha_{xy}\sigma_{xy}\right)}{\sigma^2_{xx}+\sigma^2_{xy}}\Biggr]_{r_h}=\f{U^2_0 h}{T_H T_b\Omega(B^2+h^2 A'^2_t)}+{\cal O}(k^2),\nn
\kappa_{xy}(r_h)&=&-\kappa_{yx}(r_h)=\Biggl[{\overline\kappa}_{xy}+T_H\f{\left((\alpha^2_{xx}-\alpha^2_{xy})\sigma_{xy}-2\alpha_{xx}\alpha_{xy}\sigma_{xx}\right)}{\sigma^2_{xx}+\sigma^2_{xy}}\Biggr]_{r_h}\nn&=&-\f{BU^2_0h^2(Z^2_2-A'^2_t)}{T_bT_H\Omega A'_t(B^2+Z^2_2h^2)(B^2+h^2A'^2_t)}+{\cal O}(k^2).
\eea

\subsection{Ratio of Transport quantities}

If we define the Hall angle for electrical conductivity as $\f{\sigma_{xy}}{\sigma_{xx}}$, then for the present case it reads as
\bea
\f{\sigma_{xy}}{\sigma_{xx}}&=&\bigg[ \f{B  A'_t}
{Z^2_2k^2h\psi} \bigg((B^2 T_b\Omega+k^2\psi)^2+T_b\Omega h\left[2k^2 Z^2_2\psi+T_b\Omega(B^2(Z^2_2+A'^2_t)+Z^2_2h^2A'^2_t)\right] \bigg)\Biggr]_{r_h}\nn
\eea

Similarly, for the thermoelectric case
\be
\f{\alpha_{xy}}{\alpha_{xx}}=\f{B \bigg(k^2 \psi+T_b\Omega(B^2+h^2 A'^2_t)\bigg)}{k^2  h \psi A'_t}.
\ee

\paragraph{Wiedemann-Franz relation:} It says the ratio between the thermal conductivity with the temperature times the electrical conductivity is a constant for  metal. Let us calculate such a ratio for the present case and reads as
\bea
\f{\kappa_{xx}}{T_H \sigma_{xx}}=\left[\f{{\overline\kappa}_{xx}}{T_H \sigma_{xx}}- \f{\left((\alpha^2_{xx}-\alpha^2_{xy})\sigma_{xx}+2\alpha_{xx}\alpha_{xy}\sigma_{xy}\right)}{\sigma_{xx}(\sigma^2_{xx}+\sigma^2_{xy})}\right]_{r_h}.
\eea

The precise expression of such a ratio is not that very illuminating. However, in the zero momentum dissipation limit, it shows the singular behavior
\be
Lim_{k^2\ra0}~~~\f{\kappa_{xx}}{T_H \sigma_{xx}}=\f{16 \pi^2B^2 }{ T_b\Omega\psi(B^2+h^2A'^2_t)}\f{1}{k^2}+\f{16 \pi^2B^4h^2 (Z^2_2-A'^2_t)^2}{ T^2_b\Omega^2A'^2_t(B^2+h^2A'^2_t)^2(B^2+h^2Z^2_2)^2}+{\cal O}(k^2).
\ee

\paragraph{$\alpha_{xy}/B$:} It is reported in \cite{Hartnoll:2007ih} that for cuprates the ratio $\alpha_{xy}/B$ in the small magnetic field limit goes as inverse fourth power of temperature, $\f{\alpha_{xy}}{B}(B\ra 0)\sim 1/T^4_H$. In our case, it goes as 
\be
Lim_{B\ra 0}~~~\alpha_{xy}/B\simeq \f{4\pi T_b\Omega  h}{ k^4 \psi^2}(k^2\psi+T_b\Omega h^2A'^2_t)+{\cal O}(B)^3
\ee

\paragraph{Nernst Coefficient:} It is defined as the ratio of the electric field to the thermal gradient times the magnetic field, following \cite{Hartnoll:2007ih} it reads  as 
\be
\nu=\f{1}{B}\left(\f{\alpha_{xy}\sigma_{xx}-\alpha_{xx}\sigma_{xy}}{\sigma^2_{xx}+\sigma^2_{xy}} \right).
\ee
This in the zero dissipation limit becomes
\be
Lim_{k^2\ra 0}~~~ \nu\simeq \f{4\pi h^2\psi Z^2_2(Z^2_2-A'^2_t)}{T^2_b\Omega^2A'^2_t(B^2+h^2Z^2_2)^2(B^2+h^2A'^2_t)}k^2+{\cal O}(k^3).
\ee

\subsection{In terms of charge density, $\rho$}

Let us rewrite the expressions of the conductivities in terms of the charge density, $\rho$, instead of  $A'_t$ that appear in eq(\ref{electrical_heat_current}),   and it takes the following form
\bea\label{intermsof_rho}
&&\sigma_{xx}=\sigma_{yy}=k^2T_b\Omega h\psi\left[\f{k^2\rho^2\psi+(B^2+h^2Z^2_2)(T_b\Omega\rho^2+T_b\Omega B^2Z^2_1+k^2Z^2_1\psi)}{B^2T^2_b\rho^2\Omega^2h^2Z^2_2+(\rho^2+B^2Z^2_1+h^2Z^2_1Z^2_2)(B^2T_b\Omega+k^2\psi)^2}\right],\nn
&&\sigma_{xy}=-\sigma_{yx}=\f{BT_b\Omega\rho }{Z_2\sqrt{\rho^2+Z^2_1(B^2+h^2Z^2_2)}}\times\nn
&&\left[\f{1}{B^2T^2_b\rho^2\Omega^2h^2Z^2_2+(\rho^2+B^2Z^2_1+h^2Z^2_1Z^2_2)(B^2T_b\Omega+k^2\psi)^2}\right]\times \nn 
&&\Bigg[T^2_b\rho^2\Omega^2h^2Z^2_2(B^2+h^2Z^2_2)+T_bZ^2_2\Omega h^2(\rho^2+B^2Z^2_1+h^2Z^2_1Z^2_2)(B^2T_b\Omega+2k^2\psi)+\nn&&(\rho^2+B^2Z^2_1+h^2Z^2_1Z^2_2)(B^2T_b\Omega+k^2\psi)^2\Bigg],\nn
&&T_H\alpha_{xx}=T_H\alpha_{yy}=\f{k^2T_b U_0\rho\Omega h^2Z_2\psi\sqrt{\rho^2+Z^2_1(B^2+h^2Z^2_2)}}{B^2T^2_b\rho^2\Omega^2h^2Z^2_2+(\rho^2+B^2Z^2_1+h^2Z^2_1Z^2_2)(B^2T_b\Omega+k^2\psi)^2},\nn
&&T_H\alpha_{xy}=-T_H\alpha_{yx}=\f{BT_bU_0\Omega h}{B^2T^2_b\rho^2\Omega^2h^2Z^2_2+(\rho^2+B^2Z^2_1+h^2Z^2_1Z^2_2)(B^2T_b\Omega+k^2\psi)^2}\times\nn
&&\left[Z^2_1(B^2+h^2Z^2_2)(B^2T_b\Omega+k^2\psi)+
\rho^2\left(T_b\Omega[B^2+h^2Z^2_2]+k^2\psi\right) \right],\nn
&&T_H{\overline{\kappa}}_{xx}=T_H{\overline{\kappa}}_{yy}=\f{U^2_0h\left[\rho^2+Z^2_1(B^2+h^2Z^2_2)\right](B^2T_b\Omega+k^2\psi)}{B^2T^2_b\rho^2\Omega^2h^2Z^2_2+(\rho^2+B^2Z^2_1+h^2Z^2_1Z^2_2)(B^2T_b\Omega+k^2\psi)^2},\nn
&&T_H{\overline{\kappa}}_{xy}=-T_H{\overline{\kappa}}_{yx}=
\f{BT_bU^2_0h^2\rho\Omega Z_2\sqrt{\rho^2+Z^2_1(B^2+h^2Z^2_2)}}{B^2T^2_b\rho^2\Omega^2h^2Z^2_2+(\rho^2+B^2Z^2_1+h^2Z^2_1Z^2_2)(B^2T_b\Omega+k^2\psi)^2},
\eea
where $\Omega=\f{Z_2}{(B^2+h^2Z^2_2)}\sqrt{\rho^2+Z^2_1(B^2+h^2Z^2_2)}$. All these expressions need to be evaluated at the horizon, $r_h$. It is easy to notice that the transport quantities depend on nine quantities. They are metric component, $h(r)$, couplings, $Z_1, Z_2, \psi$ and parameters, $T_b, \rho, B, k, T_H=\f{U_0}{4\pi}$.

\paragraph{Thermal conductivity:} Let us recalculate  the thermal conductivity, $\kappa_{ij}$,  in terms of the charge density explicitly. In which case the result  reads as
\bea\label{kappa_xx}
\kappa_{xx}(r_h)&=&\f{16\pi^2 T_H h\left[T_b \rho^2 Z_2\sqrt{\rho^2+Z^2_1(B^2+h^2 Z^2_2)}+k^2\psi(\rho^2+h^2Z^2_1Z^2_2)\right]}{ [2k^2 T_b\rho^2 Z_2\psi\sqrt{\rho^2+Z^2_1(B^2+h^2 Z^2_2)}+T^2_b\rho^2Z^2_2 (\rho^2+B^2 Z^2_1)+k^4\psi^2(\rho^2+h^2Z^2_1Z^2_2)]}\nn
\eea
The right hand side need to be evaluated at the horizon.
Let us calculate another quantity as defined in \cite{Blake:2017qgd}, $\kappa_L\equiv{\overline\kappa}_{xx}-T_H\alpha^2_{xx}/\sigma_{xx}$. This quantity, $\kappa_L$,  can be obtained from the thermal conductivity, $\kappa_{xx}$, in the vanishing limit of  transverse electrical conductivity and the transverse thermoelectric  conductivity, in which case, it takes the following form
\be\label{kappa_L}
\kappa_L(r_h)=\left[\f{U^2_0 h\sqrt{\rho^2+Z^2_1(B^2+h^2 Z^2_2)}}{T_H[T_bZ_2(\rho^2+B^2Z^2_1)+k^2\psi\sqrt{\rho^2+Z^2_1(B^2+h^2 Z^2_2)}]}\right]_{r_h}.
\ee
Upon expanding it in the limit of small   momentum dissipation, we get
\bea
\kappa_L(r_h)&=&\f{U^2_0 h\sqrt{\rho^2+Z^2_1(B^2+h^2 Z^2_2)}}{T_b T_H Z_2 (\rho^2+B^2 Z^2_1)}-k^2\f{U^2_0 h\psi\Bigg[\rho^2+Z^2_1 (B^2+h^2Z^2_2) \Bigg]}{ T^2_b T_H Z^2_2(\rho^2+B^2Z^2_1)^2}+ {\cal O}(k^4).
\eea
Once again the right hand side need to be evaluated at the horizon.
It is easy to see that $\kappa_{xx}$ and $\kappa_L$  matches only for vanishing momentum dissipation. 

\paragraph{At the horizon: } It is easy to derive the   differential equation  obeyed by   $U''(r)$. This    follows from eq(\ref{background_sol}) with the dependence on  the charge density and magnetic field as
\be\label{u_dd}
U''(r)-\f{T_bZ_2(\rho^2+B^2Z^2_1)}{h\sqrt{\rho^2+Z^2_1(B^2+h^2Z^2_2)}}+\f{\left[Uh^2\phi'^2-2k^2 h \psi-Uh'^2\right]}{2h^2}=0.
\ee
At the horizon the exact form of  $\kappa_L$ as written in 
eq(\ref{kappa_L}) can be re-written in terms of the function $U''(r)$ at the horizon as
\be
\kappa_L=\f{U^2_0}{T_H U''(r_h)}=\f{4\pi U_0}{U''(r_h)}=\f{4\pi U'(r_h)}{U''(r_h)}.
\ee
In getting the second equality, we have used $U_0=4\pi T_H$. This precisely matches with that obtained in \cite{Blake:2017qgd}.

\paragraph{Lim$_{k^2 \ra 0}~\kappa_{xx}$:} In the limit of vanishing momentum  dissipation, the longitudinal  component of the thermal conductivity, $\kappa_{xx}(r_h)$ can be expressed completely in terms of the metric components evaluated at the horizon in that limit. 
From eq(\ref{u_dd}), it follows in the vanishing momentum dissipation limit at the horizon
\be
U''(r_h,~k^2=0)=\left[\f{T_bZ_2(\rho^2+B^2Z^2_1)}{h\sqrt{\rho^2+Z^2_1(B^2+h^2Z^2_2)}}\right]_{r_h}
\ee
Then it follows 
\bea\label{thermal_cond_horizon_zero_k}
lim_{k^2 \ra 0}~\kappa_{xx}(r_h)&\equiv& \kappa_{xx}(r_h,~k^2=0)=\left[\f{U^2_0 h\sqrt{\rho^2+Z^2_1(B^2+h^2 Z^2_2)}}{T_b T_H Z_2 (\rho^2+B^2 Z^2_1)}\right]_{r_h}\nn &=&\f{U^2_0}{T_H U''(r_h,~k^2=0)}=\f{4\pi U'(r_h)}{U''(r_h,~k^2=0)},
\eea
where the function, $U''(r_h,~k^2=0)$, means $U''(r)$ need to be evaluated at the horizon in the limit of vanishing momentum dissipation.

\begin{figure}[h!]
	\centering
	{\includegraphics[ width=10cm,height=8cm]{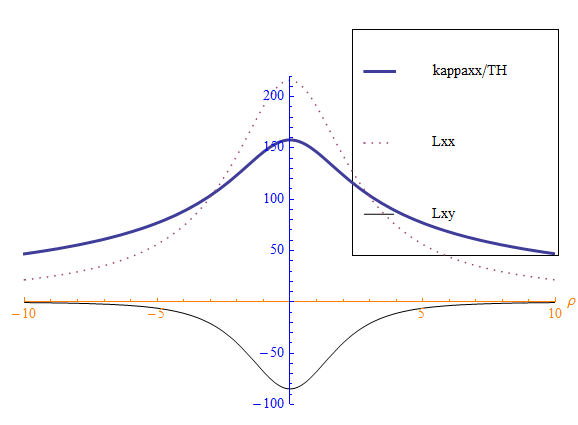}}
	\caption{
		The  figure is   plotted for $\f{\kappa_{xx}}{T_H},~~L_{xx}\equiv\f{\kappa_{xx}}{T_H\sigma_{xx}}$ and $L_{xy}\equiv\f{\kappa_{xy}}{T_H\sigma_{xy}}$ versus the charge density, $\rho$.  Moreover, the  parameters are fixed as $B=k=2=r_h$ and $Z_1=Z_2=T_b=\psi=1$ with $h(r_h)=r^2_h$. }
	\label{fig_1}
\end{figure}

\paragraph{Wiedmann-Franz relation:} The explicit form of the ratio  $L_{xx}\equiv\f{\kappa_{xx}}{T_H \sigma_{xx}}$ takes a very complicated form and is very difficult to draw any conclusion from it. The graph for  $\f{\kappa_{xx}}{T_H},~~L_{xx}\equiv\f{\kappa_{xx}}{T_H\sigma_{xx}}$ and $L_{xy}\equiv\f{\kappa_{xy}}{T_H\sigma_{xy}}$ versus  charge density, $\rho$, by fixing other parameters is plotted in fig(\ref{fig_1}). It is easy to notice that both $\f{\kappa_{xx}}{T_H},~~L_{xx}\equiv\f{\kappa_{xx}}{T_H\sigma_{xx}}$ are positive for both positive and negative charge density. It means it is an even function of charge density, which also follows from eq(\ref{intermsof_rho}) and  eq(\ref{kappa_xx}). A similar looking graph of $L_{xx}$ for Einstein-Maxwell-dilaton-axion system is reported  in \cite{Seo:2016vks}.

The form of transverse thermal conductivity takes the following form:

\bea\label{kappa_xy}
\kappa_{xy}(r_h)&=&-\f{16 \pi^2 T_H B T_b \rho h^2Z^2_1Z^2_2}{ 2k^2 T_b\rho^2 Z_2\psi\sqrt{\rho^2+Z^2_1(B^2+h^2 Z^2_2)}+T^2_b\rho^2Z^2_2 (\rho^2+B^2 Z^2_1)+k^4\psi^2(\rho^2+h^2Z^2_1Z^2_2)}.\nn
\eea
Hence, it follows that the quantity $L_{xy}\equiv\f{\kappa_{xy}}{T_H\sigma_{xy}}$ is an even function in charge density but takes negative  values and is  plotted in fig(\ref{fig_1}).

\section{Sketch of the transport quantities}

Let us re-express the transport quantities using a different set of parameters. The new set of parameters, $(B_1,~h_1,~\Omega_1,~k_1)$,  are related to the old set of parameters,  $(B,~h,~\Omega,~k)$,  in the following way 
\bea
B=\frac{\rho}{Z_1}B_1,\quad h=\frac{\rho}{Z_1Z_2}h_1, \quad k=\sqrt{\frac{T_b\rho Z_2\Omega_1}{\psi}}k_1,\quad \Omega=\frac{Z^2_!Z_2}{\rho}\Omega_1,\quad \Omega_1=\frac{\sqrt{1+B^2_1+h^2_1}}{B^2_!+h^2_1}.
\eea

With such redefinition, the background charge density, $\rho$, can be completely absorbed in the new set of parameters in such a way that it does not appear explicitly in   the transport quantities\footnote{For zero charge density, we need to use eq(\ref{intermsof_rho}) so as to understand its effect on the transport quantities.} like electrical conductivity, thermoelectric conductivity and the thermal conductivity. Moreover, with our choice of $\lambda=1$, the new set of  parameters are dimensionless. In which case,  the transport quantities reads at the horizon as
\bea
{\tilde\sigma}_{xx}&\equiv&\frac{\sigma_{xx}}{T_b Z_1}=k^2_1h_1\Omega_1\left[\frac{k^2_1+(B^2_1+h^2_1)(1+B^2_1+k^2_1) }{B^2_1h^2_1+(1+B^2_1+h^2_1)(B^2_1+k^2_1)^2} \right],\nn 
{\tilde\sigma}_{xy}&\equiv&\frac{\sigma_{xy}}{T_b Z_1}\nn&=&\frac{B_1\Omega_1}{\sqrt{1+B^2_1+h^2_1}}\left[\frac{h^2_1(B^2_1+h^2_1)+(1+B^2_1+h^2_1)[h^2_1(B^2_1+2k^2_1)+(B^2_1+k^2_1)^2]}{B^2_1h^2_1+(1+B^2_1+h^2_1)(B^2_1+k^2_1)^2}\right],\nn 
{\tilde\alpha}_{xx}&\equiv&\frac{\alpha_{xx}Z_2}{4\pi}=k^2_1h^2_1\left[\frac{\sqrt{1+B^2_1+h^2_1}}{B^2_1h^2_1+(1+B^2_1+h^2_1)(B^2_1+k^2_1)^2}\right],\nn 
{\tilde\alpha}_{xy}&\equiv&\frac{\alpha_{xy}Z_2}{4\pi}=B_1h_1\left[ \frac{(B^2_1+h^2_1)(B^2_1+k^2_1)+(B^2_1+h^2_1+k^2_1)}{B^2_1h^2_1+(1+B^2_1+h^2_1)(B^2_1+k^2_1)^2}\right],\nn
{\tilde{\overline\kappa}}_{xx}&\equiv&\frac{Z_1Z^2_2T_b}{16 \pi^2T_H}{\overline \kappa}_{xx}=\frac{h_1}{\Omega_1}\left[\frac{(1+B^2_1+h^2_1)(B^2_1+k^2_1)}{B^2_1h^2_1+(1+B^2_1+h^2_1)(B^2_1+k^2_1)^2}\right],\nn
{\tilde{\overline\kappa}}_{xy}&\equiv&\frac{Z_1Z^2_2T_b}{16 \pi^2T_H}{\overline \kappa}_{xy}=\frac{B_1h^2_1}{\Omega_1}\left[\frac{\sqrt{1+B^2_1+h^2_1}}{B^2_1h^2_1+(1+B^2_1+h^2_1)(B^2_1+k^2_1)^2}\right].
\eea

This gives the longitudinal and the transverse thermal conductivities as 
\bea
{\tilde \kappa}_{xx}&\equiv&\frac{Z_1Z^2_2T_b}{16 \pi^2T_H}\kappa_{xx}={\tilde{\overline\kappa}}_{xx}-
\frac{({\tilde\alpha}^2_{xx}-{\tilde\alpha}^2_{xy}){\tilde\sigma}_{xx}+2{\tilde\alpha}_{xx}{\tilde\alpha}_{xy}{\tilde\sigma}_{xy}}{{\tilde\sigma}^2_{xx}+{\tilde\sigma}^2_{xy}},\nn 
{\tilde \kappa}_{xy}&\equiv&\frac{Z_1Z^2_2T_b}{16 \pi^2T_H}\kappa_{xy}={\tilde{\overline\kappa}}_{xy}+
\frac{({\tilde\alpha}^2_{xx}-{\tilde\alpha}^2_{xy}){\tilde\sigma}_{xy}-2{\tilde\alpha}_{xx}{\tilde\alpha}_{xy}{\tilde\sigma}_{xx}}{{\tilde\sigma}^2_{xx}+{\tilde\sigma}^2_{xy}}.
\eea

\paragraph{Transport quantities:} For the computation and  sketch of the transport quantities at the horizon, the function $h_1(r_h)$ is set as $r^2_h$, which is expected for AdS spacetime with an exception for $AdS_2$ spacetime.  Note,  $r_h$ denotes dimensionless horizon only for this section.

\begin{figure}[h!]
	\centering
	{\includegraphics[ width=10cm,height=8cm]{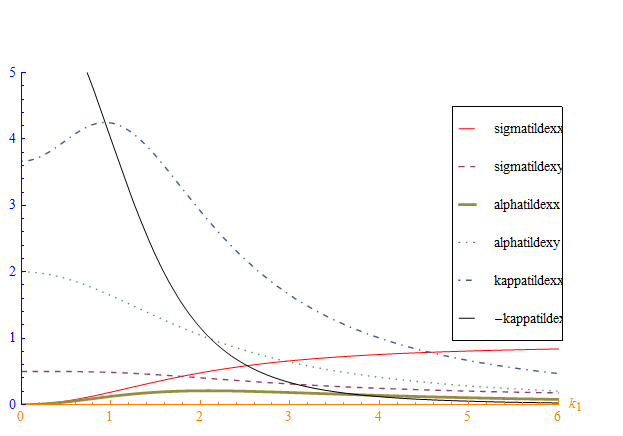}}
	\caption{
		The  figure is   plotted for the   transport quantities  versus the modified dissipation parameter, $k_1$.  Moreover, the  parameters are fixed as $B_1=2=r_h$. The notation is sigmatildexx=${\tilde\sigma}_{xx}$,~sigmatildexy=${\tilde\sigma}_{xy}$,~alphatildexx=${\tilde\alpha}_{xx}$,~alphatildexy=${\tilde\alpha}_{xy}$,~kappatildexx=${\tilde\kappa}_{xx}$, ~kappatildexy=${\tilde\kappa}_{xy}$. }
	\label{fig_2}
\end{figure} 

It follows from the figure Fig(\ref{fig_2}) that the longitudinal electrical conductivity, ${\tilde\sigma}_{xx}$,  and the longitudinal component of the thermoelectric conductivity, ${\tilde\alpha}_{xx}$, vanishes for vanishing dissipation,  for fixed $B_1$ and $r_h$.  This is mainly due to the fact that the electric current and the heat current along that direction vanishes for vanishing dissipation. However, the rest of the transport quantities remains non-zero for vanishing dissipation.

\begin{figure}[h!]
	\centering
	{\includegraphics[ width=10cm,height=8cm]{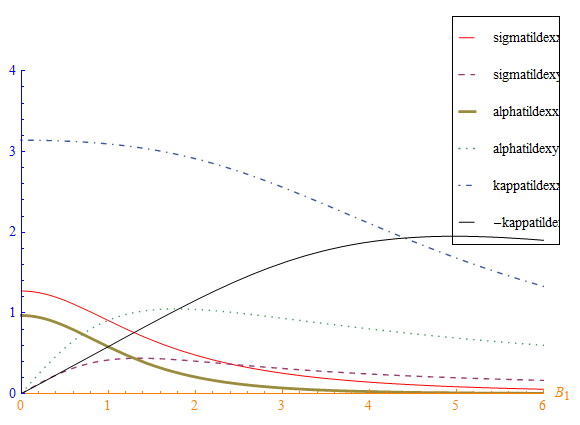}}
	\caption{
		The  figure is   plotted for the   transport quantities  versus the modified magnetic field parameter, $B_1$.  Moreover, the  parameters are fixed as $k_1=2=r_h$. The notation is sigmatildexx=${\tilde\sigma}_{xx}$,~sigmatildexy=${\tilde\sigma}_{xy}$,~alphatildexx=${\tilde\alpha}_{xx}$,~alphatildexy=${\tilde\alpha}_{xy}$,~kappatildexx=${\tilde\kappa}_{xx}$, ~kappatildexy=${\tilde\kappa}_{xy}$. }
	\label{fig_3}
\end{figure}

 Instead of fixing the modified magnetic field, $B_1$, and the  size of the horizon, $r_h$, if we fix the modified dissipation parameter, $k_1$, and the size of the horizon,  the transport quantities  are plotted in figure Fig(\ref{fig_3}). It follows that the transverse electrical conductivity (Hall conductivity), ${\tilde\sigma}_{xy}$, transverse thermoelectric conductivity, ${\tilde\alpha}_{xy}$ and the transverse  thermal conductivity ${\tilde\kappa}_{xy}$ vanishes for vanishing modified magnetic field, $B_1$. This is due to the fact that the electric current as well as the heat current vanishes along that direction for vanishing $B_1$.

\begin{figure}[h!]
 	\centering
 	{\includegraphics[ width=10cm,height=8cm]{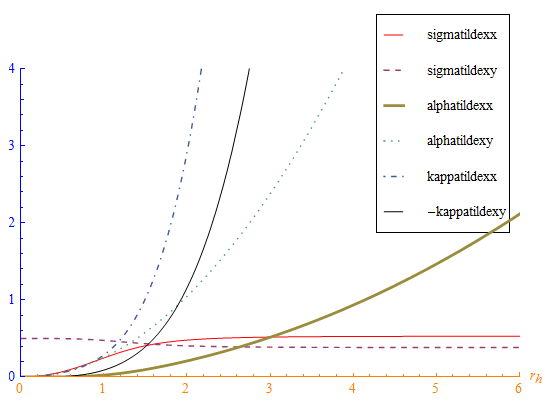}}
 	\caption{
 		The  figure is   plotted for the   transport quantities  versus the  size of the horizon, $r_h$.  Moreover, the  parameters are fixed as $k_1=2=B_1$. The notation is sigmatildexx=${\tilde\sigma}_{xx}$,~sigmatildexy=${\tilde\sigma}_{xy}$,~alphatildexx=${\tilde\alpha}_{xx}$,~alphatildexy=${\tilde\alpha}_{xy}$,~kappatildexx=${\tilde\kappa}_{xx}$, ~kappatildexy=${\tilde\kappa}_{xy}$. }
 	\label{fig_4}
 \end{figure}

If we fix the modified magnetic field, $B_1$, and the modified dissipation parameter, $k_1$, then the sketch of the transport quantities  is given in figure fig(\ref{fig_4}).

 \subsection{Transport quantities in different limits:}
 
 In what follows, the discussion will be performed using the following set of parameters,  $(B,~h,~\Omega,~k)$.
 
 \paragraph{Transport quantities for zero charge density, $\rho=0$, limit:} In this limit the transport quantities takes the following form
 \bea\label{transport_rho_0}
 \sigma^{-1}_{xx}&=&\frac{\sqrt{B^2+h^2 Z^2_2}}{Z_1Z_2T_b h}+\frac{B^2}{k^2 h \psi},\quad
 \sigma_{xy}=0,\quad \alpha_{xx}=0,\nn T^{-1}_H\alpha^{-1}_{xy}&=&\frac{B}{U_0h}+\frac{k^2\psi}{BT_b U_0\Omega h}=\frac{B}{U_0h}+\frac{k^2\psi}{BT_b U_0 h}\frac{\sqrt{B^2+h^2 Z^2_2}}{Z_1 Z_2}\nn
 T^{-1}_H\overline{\kappa}^{-1}_{xx}&=&\frac{B^2 T_b}{U^2_0 h}\frac{Z_1 Z_2}{\sqrt{B^2+h^2 Z^2_2}}+\frac{k^2\psi}{U^2_0 h},\quad \overline{\kappa}_{xy}=0.
 \eea
 It is easy to notice that there occurs two types of term in the inverse of the transport quantities. One term is independent of the dissipation parameter $k$  and the other term depends on it.  Note, the separation is possible only for the inverse of the transport quantities but not for the transport quantities. So, for the transport quantities there does not exists any clear cut separation of its  dependence on the dissipation parameter and the magnetic field.
 
 The way these terms arises can be understood as follows. From the electric currents as written down  in eq(\ref{electric_current_horizon}), it follows that there are four terms that contribute to electric currents both along x and y direction. For example, the contribution to the  electric current along x-direction in the absence of (background) charge density is due to  the electric field, $E_x$ and due to the $ty$ component of the metric fluctuation. The latter essentially arises at the horizon mainly due to an in-falling boundary condition at the horizon, which creates an electric field along x-direction and a thermal gradient along y-direction.
 
  Hence, the total electric current along x-direction is due to an electric field produced by particle-hole pair  and  another contribution of an  electric field along x-direction comes due to momentum dissipation and presence of background magnetic field, whereas the thermal gradient  contributes only along y-direction.
 
 \paragraph{Transport quantities for zero magnetic field, $B=0$, limit:} In this limit, the transport quantities take the following form:
 \bea
 \sigma_{xx}&=&\frac{T_b}{hZ_2}\sqrt{\rho^2+h^2Z^2_1Z^2_2}+\f{\rho^2}{k^2 h \psi},\quad \sigma_{xy}=0,\quad \alpha_{xx}=\frac{T_b U_0\rho}{T_H k^2\psi},\nn
 \alpha_{xy}&=&0,\quad \overline{\kappa}_{xx}=\f{U^2_0 h}{T_H k^2\psi},\quad 
 \overline{\kappa}_{xy}=0.
 \eea
 
 For the transport quantities, there does  exists a clear cut separation between the contribution due to charge conjugation symmetric term and contribution due to dissipation. The Hall conductivity vanishes for zero magnetic field. This is simply due to  non-existence of a current in the direction perpendicular to the direction of electric field. Similarly, there is no electric current in the direction perpendicular to the direction of thermal gradient. The longitudinal component of the thermoelectric conductivity and $\overline{\kappa}_{xx}$ is purely due to dissipation. 
 
 \paragraph{Transport quantities for zero momentum dissipation, $k=0$, limit:}
 In this limit, the transport quantities take the following form:
 \bea
 \sigma_{xx}&=&\frac{h \psi}{B^2}k^2+{\cal O} (k^4),\quad \sigma_{xy}=\frac{T_b \rho }{B}-{\cal O} (k^4),\quad \alpha_{xx}=\frac{ U_0\rho h^2 \psi k^2}{T_H T_b B^2(\rho^2+B^2 Z^2_1)}+{\cal O} (k^4),\nn
 \alpha_{xy}&=&\frac{U_0 h}{T_H B}+{\cal O} (k^2),\quad \overline{\kappa}_{xx}=\f{U^2_0 h\sqrt{\rho^2+Z^2_1(B^2+h^2Z^2_2)}}{T_H T_b Z_2(\rho^2+B^2Z^2_1)}+{\cal O} (k^2),\nn
 \overline{\kappa}_{xy}&=&\frac{ U^2_0\rho h^2 }{T_H T_b B(\rho^2+B^2 Z^2_1)}+{\cal O} (k^2).
 \eea
 
 In the situation for which there is no dissipation implies the absence of the electric current in the direction of the electric field as well as in the direction of the thermal gradient. It means the longitudinal component of electrical conductivity as well as the longitudinal component of the thermoelectric conductivity vanishes. However, there exists an electric current in a direction perpendicular to the direction of electric field and thermal gradient. Hence, the transverse electrical conductivity as well as the transverse thermoelectric conductivity are non-zero.

 \paragraph{Transport quantities for very high  dissipation, $k=\infty$, limit:}
 
 In the limit of very high dissipation but with finite charge density and magnetic field, the transport quantities take the following form:
 \bea
 \sigma_{xx}&=&\frac{T_b h Z_2\sqrt{\rho^2+Z^2_1(B^2+h^2Z^2_2)}}{B^2+h^2 Z^2_2}+{\cal O} (1/k^2),\quad \sigma_{xy}=\frac{T_b \rho B }{B^2+h^2 Z^2_2}+{\cal O} (1/k^2),\nn \alpha_{xx}&=&\frac{T_b U_0\rho h^2 Z_2^2}{T_H \psi (B^2+h^2 Z^2_2)k^2}+{\cal O} (1/k^4),\quad
 \alpha_{xy}=\frac{B T_b U_0 hZ_2\sqrt{\rho^2+Z^2_1(B^2+h^2Z^2_2)}}{T_H (B^2+h^2 Z^2_2)\psi k^2}+{\cal O} (1/k^4),\nn \overline{\kappa}_{xx}&=&\f{U^2_0 h}{T_H \psi k^2}+{\cal O} (1/k^4),\quad
 \overline{\kappa}_{xy}=\frac{ T_b B U^2_0\rho h^2Z^2_2 }{T_H  (B^2+h^2 Z^2_2)\psi^2 k^4}+{\cal O} (1/k^6).
 \eea
 
 This is simply due to the fact that the fluctuation of the  metric components at the horizon vanishes, in particular, it goes as $1/k^2$. Hence, the total electric current along a particular direction is due to an electric field along the same direction and an electric field in the perpendicular direction. This explains the non-vanishing behavior of the longitudinal electrical conductivity as well as the transverse  electrical conductivity. The thermal gradients does not contribute to the electric currents.

\section{Example at UV}

Let us find an exact solution of the system as written down in eq(\ref{action}) in a very specific case, i.e., for the asymptotically AdS solution with constant dilaton, non-zero axion and study its transport properties. In particular, we make the choice as 
\bea
\phi(r)&=&\phi_{0},\quad  V(\phi)=0,\quad W_1(\phi)=W_2(\phi)=1,\quad Z_1(\phi)=Z_2(\phi)=1\nn h(r)&=&\f{r^2}{R^2},\quad U(r)=\f{r^2}{R^2} f(r),\quad \chi_1=k~x,\quad \chi_2=k~y,
 \eea
  where $R$ is the AdS radius. It is easy to see that   the equation of motion for the dilaton and axion are satisfied. The equation of  motion associated to  the metric components gives one first order differential equation for $f(r)$ and one second order differential equation for $f(r)$. It reads as
  \bea
  &&r~ f''(r)+6f'(r)+\f{2R^2\Lambda}{r}+\f{6f(r)}{r}+
  T_b \f{r R^2}{\sqrt{r^4+R^4(\rho^2+B^2)}}=0,\nn
  	&&r~f'(r)+3f(r)+\Lambda R^2+\f{k^2 R^4}{2r^2}+T_bR^2 \f{\sqrt{r^4+R^4(\rho^2+B^2)}}{2r^2}=0.
  \eea
  In fact, it is easy to check that these two differential equations are not independent of each other. The solution of the differential equation reads in terms of the Hypergeometric function, ${}_2F_1\left[a,b,c,x\right]$, as
  \bea
  f(r)&=&\f{c_1}{r^3}-\f{\Lambda R^2}{3}+\f{k^2 R^4}{2r^2}-T_b R^2\f{\sqrt{r^4+R^4(\rho^2+B^2)}}{6r^2}\nn&-&T_b \f{R^4\sqrt{\rho^2+B^2}}{3r^2}{}_2F_1\left[\f{1}{2}, \f{1}{4}, 1+\f{1}{4}, -\f{r^4}{R^4(\rho^2+B^2)} \right],
  \eea
  where $c_1$ is the constant of integration.
  The effect of the momentum dissipation, $k$, appears in the metric through the function $f(r)$. Now, the gauge field has the following structure
  \be
   A'_t=\f{\rho R^2}{\sqrt{r^4+R^4(\rho^2+B^2)}}.
  \ee
  
  \paragraph{Hawking Temperature:} The Hawking temperature, $T_H$,  for such a solution reads as
  \be
  T_H=-\f{r_h}{8\pi}\left[ 2\Lambda+\f{k^2R^2}{r^2_h}+\f{T_b}{r^2_h}\sqrt{r^4_h+R^4(\rho^2+B^2)}\right].
  \ee
  
  It is generally very difficult to solve such an equation and read out the size of the horizon, $r_h$, in terms of the temperature, $T_H$. However,  
  in the low temperature limit, $T_H\ra 0$, the solution of the size of the horizon admits a Taylor series expansion like 
  \be
  r_h=r_0+ r_1~T_H+{\cal O}(T_H)^2,
  \ee
  where $r_0$ and $r_1$ are independent of temperature. In fact, there exists two different choices of $r_0$ and $r_1$, which is fixed by the value of the tension of the brane, $T_b$.
  
  \paragraph{ Choice $T^2_b\neq 4\Lambda^2$:} The choice of the quantities for this case   are
  \bea
  r_0&=&\sqrt{\f{2k^2R^2\Lambda\mp\sqrt{R^4 T^2_b\left[k^4-(T^2_b-4\Lambda^2)(\rho^2+B^2)\right]}}{T^2_b-4\Lambda^2}},\nn
  r_1&=&\left(\f{4\pi}{R^2(T^2_b-4\Lambda^2)\left[k^4-(T^2_b-4\Lambda^2)(\rho^2+B^2)\right]}\right)\times\nn
  &&\Bigg[2k^4R^2\Lambda-2R^2\Lambda(T^2_b-4\Lambda^2)
  (\rho^2+B^2)\mp k^2\sqrt{R^4 T^2_b\left[k^4-(T^2_b-4\Lambda^2)(\rho^2+B^2)\right]} \Bigg].\nn
  \eea
  
 \paragraph{ Choice $T^2_b= 4\Lambda^2$:} For this choice  the quantities  are
  \be
  r_0=\mp\f{R}{2k}\sqrt{\f{T^4_b(\rho^2+B^2)-k^4}{\Lambda}},\quad 
  r_1=-\f{\pi}{k^4\Lambda}\left[T^4_b(\rho^2+B^2)+k^4 \right].
  \ee
 
 It is interesting to note that for a such choice of the tension of brane,  $T^2_b= 4\Lambda^2$, one cannot take the zero dissipation limit, $k^2\neq 0$.
 
 \subsection{In the low temperature limit: $T_H\ra 0$}
 We shall write the explicit structure of the transport quantities in the zero temperature limit. In fact, 
 it is not easy and very elegant to see the  full temperature dependence of the transport quantities. So, we shall be looking at the zero temperature limit of it.

\be
\sigma_{xx}=\sigma_{yy}= 
\left\{
\begin{array}{l l}
	\f{r^2_0}{B^2R^2} k^2+2\f{r_0 r_1}{B^2R^2}k^2 T_H+{\cal} O(k^4,~T^2_H)
 & \quad \textrm{in the $k^2\ra 0$ limit}\\
\f{r^2_0 T_b\sqrt{R^4(\rho^2+B^2)+r^4_0}}{B^2R^4+r^4_0}+
&\\2\f{R^4r_0r_1T_b[B^4R^4-r^4_0\rho^2+B^2(r^4_0+R^4\rho^2)]}{(B^2R^4+r^4_0)^2\sqrt{R^4(\rho^2+B^2)+r^4_0}}T_H+ {\cal O} (1/k^2,~ T^2_H) & \\
\quad\quad\quad&\quad \textrm{In the  $k^2\ra \infty$ limit}.\\
\end{array} \right.
\ee

 \be
 \sigma_{xy}=-\sigma_{yx}= 
 \left\{
 \begin{array}{l l}
 	\f{T_b\rho}{B}-\f{r^4_0\rho k^4}{B^3 R^4T_b(B^2+\rho^2)} -4\f{r^3_0 r_1\rho k^4}{B^3R^4T_b(\rho^2+B^2)} T_H+{\cal} O(k^6,~T^2_H)
 	& \quad \textrm{in the $k^2\ra 0$ limit}\\
 	&\\\f{BR^4T_b\rho}{(B^2R^4+r^4_0)}-4\f{BR^4r^3_0r_1T_b\rho}{(B^2R^4+r^4_0)^2}T_H+ {\cal O} (1/k^2,~ T^2_H) & \\
 	\quad\quad\quad&\quad \textrm{In the  $k^2\ra \infty$ limit}.\\
 \end{array} \right.
 \ee

 \be
 \alpha_{xx}=\alpha_{yy}= 
 \left\{
 \begin{array}{l l}
 	\f{4\pi r^4_0\rho k^2}{B^2R^4T_b(B^2+\rho^2)} +\f{16\pi r^3_0 r_1\rho k^2}{B^2R^4T_b(B^2+\rho^2)} T_H+{\cal} O(k^4,~T^2_H)
 	& \quad \textrm{in the $k^2\ra 0$ limit}\\
 	&\\\f{4\pi r^4_0\rho T_b}{(B^2R^4+r^4_0)k^2}+\f{16\pi B^2 R^4 r^3_0r_1\rho T_b}{(B^2R^4+r^4_0)^2k^2}T_H+ {\cal O} (1/k^4,~ T^2_H) 
 	\quad\quad\quad&\quad \textrm{In the  $k^2\ra \infty$ limit}.\\
 \end{array} \right.
 \ee

 \be
 \alpha_{xy}=-\alpha_{yx}= 
 \left\{
 \begin{array}{l l}
 	\f{4\pi r^2_0}{BR^2} +\f{8\pi r_0 r_1}{BR^2} T_H+{\cal} O(k^2,~T^2_H)
 	& \quad \textrm{in the $k^2\ra 0$ limit}\\
 	&\\\f{4\pi B  r^2_0 T_b\sqrt{r^4_0+R^4(\rho^2+B^2)}}{(B^2R^4+r^4_0)k^2}+\\
 	&\\\f{8\pi B R^4 r_0r_1T_b(B^2R^4-r^4_0\rho^2+B^2(r^4_0+R^4\rho^2))}{(B^2R^4+r^4_0)^2k^2\sqrt{r^4_0+R^4(\rho^2+B^2)}}T_H+ {\cal O} (1/k^4,~ T^2_H) & \\
 	\quad\quad\quad&\quad \textrm{In the  $k^2\ra \infty$ limit}.\\
 \end{array} \right.
 \ee

 \be
 {\overline{\kappa}}_{xx}={\overline{\kappa}}_{yy}= 
 \left\{
 \begin{array}{l l}
 	\f{16\pi^2 r^2_0\sqrt{r^4_0+R^4(\rho^2+B^2)}}{R^4T_b(B^2+\rho^2)}T_H +{\cal} O(k^2,~T_H)
 	& \quad \textrm{in the $k^2\ra 0$ limit}\\
 	&\\\f{16\pi^2 r^2_0}{R^2k^2}T_H+ {\cal O} (1/k^4,~ T^2_H) 
 	\quad\quad\quad&\quad \textrm{In the  $k^2\ra \infty$ limit}.\\
 \end{array} \right.
 \ee
 
 \be
 {\overline{\kappa}}_{xy}=-{\overline{\kappa}}_{yx}= 
 \left\{
 \begin{array}{l l}
 	\f{16\pi^2 r^4_0\rho}{BR^4T_b(B^2+\rho^2)}T_H +{\cal} O(k^2,~T^2_H)
 	& \quad \textrm{in the $k^2\ra 0$ limit}\\
 	&\\\f{16\pi^2 B \rho T_b r^4_0}{(B^2R^4+r^4_0)k^4}T_H+ {\cal O} (1/k^6,~ T^2_H) 
 	\quad\quad\quad&\quad \textrm{In the  $k^2\ra \infty$ limit}.\\
 \end{array} \right.
 \ee

 One of the interesting observation is that in the absence of any charge density, $\rho\ra 0$,  the off diagonal component of the electrical and thermal conductivity vanishes whereas the thermoelectric conductivity does not vanish, which is in agreement with the result shown in eq(\ref{transport_rho_0}).
 
 From the result of the computation, it follows that  the transport quantities can be written  in the low temperature limit for $AdS_4$ solution as follows:
 \bea
 \sigma_{xx}&=&\sigma_{yy}=\sigma_0+\sigma_1 T_H,\quad \sigma_{xy}=-\sigma_{yx}=\sigma_2+\sigma_3 T_H,\nn
 \alpha_{xx}&=&\alpha_{yy}=\alpha_0+\alpha_1 T_H,\quad 
  \alpha_{xy}=-\alpha_{yx}=\alpha_2+\alpha_3 T_H,\nn
  \kappa_{xx}&=&\kappa_{yy}=\kappa_0 T_H,\quad  \kappa_{xy}=-\kappa_{yx}=\kappa_1 T_H,
 \eea
 where $\sigma_i,~\alpha_i,~\kappa_i$ are independent of temperature. Given the results for electrical, thermoelectric and thermal conductivities at low temperature, the Wiedemann-Franz relation takes the following  form  
 \bea
 Lim_{T_H\ra 0}~~  \frac{\kappa_{xx}}{T_H \sigma_{xx}}=~\frac{\kappa_0}{\sigma_0},\quad 
 Lim_{T_H\ra 0}~~  \frac{\kappa_{xy}}{T_H \sigma_{xy}}=~\frac{\kappa_1}{\sigma_2}.
 \eea
 
 \section{ Example at IR}
 
 Let us study another example but at IR. In order to construct such a solution at IR, namely, $AdS_2\times R^2$, the choice of the functions are 
 \be
 Z_1(\phi)=z_1,\quad Z_2(\phi)=z_2,\quad h(r)=L^2,\quad V(\phi)=0,\quad  \phi(r)=0,\quad \psi(r)=\psi_0,
 \ee
 where $z_1,~z_2,~\psi_0$ and $L^2$ are constants. The function 
 \be
  U(r)=\f{r^2}{R^2_2}\left(1-\f{r_h}{r}\right),\quad F=\f{\rho z_2}{\sqrt{\rho^2+z^2_1(B^2+L^4 z^2_2)}}dr\w dt+B dx\w dy.
 \ee
 
 In which case, it is easy to see that the equation of motion of the scalar field eq(\ref{scalar_eom}) gets satisfied automatically. The equation of motion of  the metric components eq(\ref{background_sol}) gives
 \bea 
 &&2\Lambda L^2+k^2\psi_0+T_b z_2\sqrt{\rho^2+z^2_1(B^2+L^4 z^2_2)}=0,\nn
 &&2\Lambda R^2_2+2 +T_bL^2 R^2_2 \f{ z^2_1 z^3_2}{\sqrt{\rho^2+z^2_1(B^2+L^4 z^2_2)}}=0.
 \eea
 
 The solution is
 \be
 \Lambda=-\f{1}{R^2_2}-T_b\f{L^2 z^2_1 z^3_2}{2\sqrt{\rho^2+z^2_1(B^2+L^4 z^2_2)}},\quad \psi_0=\f{2L^2}{k^2R^2_2}-T_b\f{z_2(\rho^2+B^2 z^2_1)}{k^2\sqrt{\rho^2+z^2_1(B^2+L^4 z^2_2)}}.
 \ee
 The Hawking temperature for the present case reads as $T_H=r_h/(4\pi R^2_2)$. Moreover, 
 the electrical and thermal transport quantities behave in the following manner
 \be\label{electric_thermo_cond_ads2}
 \sigma_{xx}={\rm constant},\quad \sigma_{xy}={\rm constant},\quad \alpha_{xx}={\rm constant},\quad \alpha_{xy}={\rm constant},\quad {\overline{\kappa}}_{xx}\sim T_H,\quad {\overline{\kappa}}_{xy}\sim T_H.
 \ee
 Here constant means they are independent of temperature but depends on charge density and magnetic field.
 So, the thermal conductivity  goes linear with temperature, $\kappa_{ij}\sim T_H$. It just follows that 
 \bea\label{thermal_cond_ads2}
  \frac{\kappa_{xx}}{T_H \sigma_{xx}}=~{\rm constant},\quad 
   \frac{\kappa_{xy}}{T_H \sigma_{xy}}=~{\rm constant}.
 \eea
 It means  the ratio of the longitudinal(transverse) component of the thermal conductivity with the temperature times the longitudinal(transverse) electrical conductivity respects\footnote{It means, it is independent of temperature but does depends on the  charge density and magnetic field.} the Wiedemann-Franz relation.  
 
\begin{figure}[h!]
 	\centering
 	{\includegraphics[ width=10cm,height=8cm]{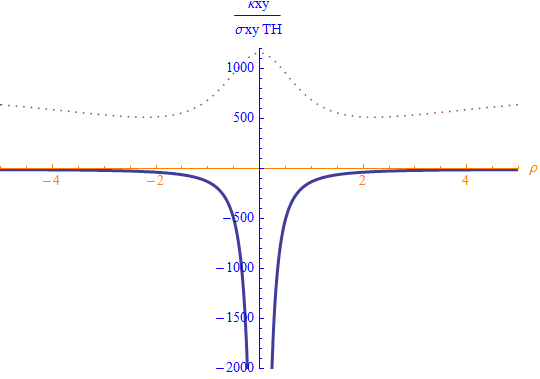}}
 	\caption{
 		The  figure is   plotted for  $L_{xy}\equiv\f{\kappa_{xy}}{T_H\sigma_{xy}}$ versus the charge density, $\rho$.  Moreover, the  parameters are fixed as $B=2$ and $L=z_1=z_2=T_b=1$. The parameter   $R_2=1$ for the thick curve whereas $R_2=2$ for the dotted curve. }
 	\label{fig_5}
 \end{figure}

 In order to have a feeling of  the Hall-Lorentz ratio at IR i.e., with respect to the $AdS_2\times R^2$ solution, we have plotted it  in Fig(\ref{fig_5}). It follows from the figure that the Hall-Lorentz ratio can be both positive and negative.
 
\begin{figure}[h!]
 	\centering
 	{\includegraphics[ width=10cm,height=8cm]{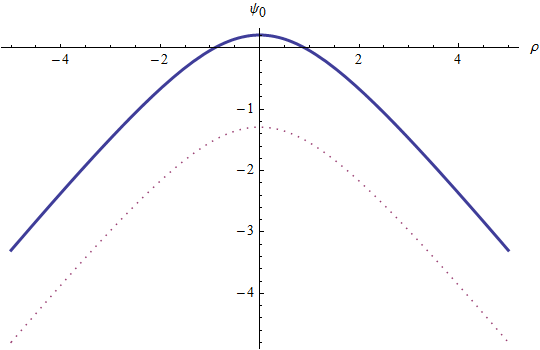}}
 	\caption{
 		The  figure is   plotted for  $\psi_0$ versus the charge density, $\rho$.  Moreover, the  parameters are fixed as $B=2$ and $L=z_1=z_2=T_b=1$. The parameter   $R_2=1$ for the thick curve whereas $R_2=2$ for the dotted curve. }
 	\label{fig_6}
 \end{figure}

Let us note that depending on the choice of parameter, $\psi_0$,  can take both positive and negative values. In particular, for the choice of parameters $B=2$ and $L=z_1=z_2=T_b=1$, $\psi_0$ is  positive for $R_2=1$, whereas negative  for $R_2=2$ for some values of charge density, which is plotted in Fig(\ref{fig_6}). 
 
 \paragraph{An observation:} Instead of computing the quantities like $\frac{\kappa_{xx}}{T_H \sigma_{xx}}$ and $\frac{\kappa_{xy}}{T_H \sigma_{xy}}$,  if we compute the matrix $\frac{1}{T_H}\kappa\cdot \sigma^{-1}$  then its components are constant too.
 \bea
\frac{1}{T_H}\kappa\cdot \sigma^{-1}&=&\frac{1}{T_H(\sigma^2_{xx}+\sigma^2_{xy})}
 \left(\begin{array}{cc} \kappa_{xx}\sigma_{xx}+\kappa_{xy}\sigma_{xy} & -\kappa_{xx}\sigma_{xy}+\kappa_{xy}\sigma_{xx}\\ \kappa_{xx}\sigma_{xy}-\kappa_{xy}\sigma_{xx} & \kappa_{xx}\sigma_{xx}+\kappa_{xy}\sigma_{xy} \end{array}\right)\nn
 &=&{\rm constant},
 \eea
 
 where we have used eq(\ref{electric_thermo_cond_ads2}) and eq(\ref{thermal_cond_ads2}). Here constant means it is independent of temperature but does depends on the  charge density and magnetic field.

\section{Conclusion}

It follows from the study of \cite{Donos:2014cya} that in  the absence of the magnetic field the perturbation of the metric component breaks down in the zero momentum dissipation limit. In particular,  for vanishing magnetic field the metric component $h_{tx}$ diverges\footnote{See eq(3.6) of \cite{Donos:2014cya}.} as the momentum dissipation vanishes, $k^2\ra 0$, in which case, the perturbation of the metric component $h_{tx}$ breaks down. So the zero momentum dissipation limit is very subtle to consider as far as the perturbations are concerned. 
Here we show that in the presence of a magnetic field, the perturbation does not breakdown. A feature  persists in the context of the computation of the transport quantities in Einstein-Maxwell system in \cite{Blake:2014yla}, \cite{Blake:2015ina} and \cite{Kim:2015wba}.

In our study, we find 
there exists an important but subtle difference between the electric currents and the heat currents at the horizon. It simply follows from the fact that  the  electric currents at the horizon depends explicitly on three different quantities (a) on the background charge density, $\rho$ (b) on the fluctuation of the metric components, $h_{tx}$ and $h_{ty}$ at the horizon (c) on the applied electric fields, ${\vec E}$, and magnetic field, ${\vec B}$. The  heat currents at the horizon depends only on the fluctuation of the metric components, $h_{tx}$ and $h_{ty}$. However, due to in-falling boundary condition at the horizon, the metric fluctuations $h_{tx}$ and $h_{ty}$ at the horizon  in turn depends on the charge density and  applied electric fields, magnetic field and thermal gradients. 

We have computed the transport quantities for Einstein-DBI system. The results shows that the longitudinal part of the conductivities   matches precisely with that obtained in \cite{Blake:2017qgd}. The transverse part of the electrical conductivity matches with that reported in \cite{Cremonini:2017qwq}. The transverse part of the thermoelectric conductivity as well as the thermal conductivity are new results.

We observed that the longitudinal component of the thermal conductivity at the horizon in the zero momentum dissipation limit can be expressed completely in terms of the metric component in that limit. In \cite{Blake:2017qgd}, it is shown that the longitudinal thermal conductivity at the horizon in a specific case, namely,  for the vanishing transverse electrical conductivity as well as the transverse thermoelectric conductivity, can be expressed  in terms of the metric component at the horizon but with any value of momentum dissipation. It follows that the longitudinal thermal  conductivity is positive, whereas, the transverse  thermal  conductivity takes negative values for positive values of charge density. The Lorentz ratio,  $L_{xx}=\frac{\kappa_{xx}}{T_H \sigma_{xx}}$, takes positive values and the graph looks similar  in form to that obtained for the Einstein-Maxwell system in \cite{Seo:2016vks}. The Hall Lorentz ratio,  $L_{xy}=\frac{\kappa_{xy}}{T_H \sigma_{xy}}$, takes negative values.

The exact temperature dependence of the transport quantities are obtained by studying two explicit examples, one at UV and the other at IR.
The result of our study in the low temperature limit  at UV and IR can be summarized in the table(\ref{table}). The precise momentum dissipation dependence for the $AdS_4$  case is suppressed, for simplicity. 
\begin{table}
\begin{center}
	\begin{tabular}{ |l | p{3cm} | p{2cm} |}
		\hline
		Transport quantities in Einstein-DBI system  & UV  & IR \\ 
		& ($AdS_4$) & ($AdS_2\times R^2$)\\ \hline\hline
		Longitudinal electrical conductivity, $\sigma_{xx}=\sigma_{yy}$ &$\sigma_0+\sigma_1 T_H$& constant\\\hline
		Transverse electrical conductivity, $\sigma_{xy}=-\sigma_{yx}$ &$\sigma_2+\sigma_3 T_H$ &constant\\\hline
Longitudinal thermoelectric conductivity, $\alpha_{xx}=\alpha_{yy}$ &$\alpha_0+\alpha_1 T_H$&constant\\\hline
Transverse thermoelectric conductivity, $\alpha_{xy}=-\alpha_{yx}$ &$\alpha_2+\alpha_3 T_H$&constant\\\hline
$Lim_{B\ra 0,~T_H\ra 0}~~ \f{\alpha_{xy}}{B}$&constant&constant\\ \hline
Longitudinal thermal conductivity, $\kappa_{xx}=\kappa_{yy}$ &$\kappa_0 T_H$& $T_H$\\\hline
Transverse thermal conductivity, $\kappa_{xy}=-\kappa_{yx}$ &$\kappa_1 T_H$& $T_H$\\\hline\hline
$Lim_{T_H\ra 0}~~  \frac{\kappa_{xx}}{T_H \sigma_{xx}}$&constant&constant\\\hline
		$ Lim_{T_H\ra 0}~~  \frac{\kappa_{xy}}{T_H \sigma_{xy}}$&constant&${\rm constant}$\\\hline\hline
		Components of $\left(Lim_{T_H\ra 0}~~ \frac{\kappa\cdot \sigma^{-1}}{T_H}\right)$&constant&constant\\\hline\hline
	\end{tabular}
\end{center}\caption{\label{table}The transport  quantities at low temperature, with  $\sigma_0,~\sigma_1,~ \alpha_0,~ \alpha_1,~\kappa_0$ and $\kappa_1$ are constants. It means these are independent of temperature but depends on dissipation parameter, charge density and magnetic field.}
\end{table}

It is suggested in \cite{Hartnoll:2007ih} that in the presence of a background magnetic field, a proper subtraction of the magnetization   current and the energy magnetization need to be done in heat current, which we leave for future study.

\paragraph{Acknowledgment:} It is a pleasure to thank the anonymous referee for many constructive suggestions and questions.  My thanks are to  arxiv.org, gmail of google.com and inspirehep.net for providing their support through out this work. I am thankful to Professor Ken Intriligator for his kindness to let me know  the error while submitting it to arxiv.

\end{document}